\RequirePackage{ifpdf}
\ifpdf 
\documentclass[pdftex]{sigma}
\else
\documentclass{sigma}
\fi

\newcommand{\pard}{[\! [ }
\newcommand{\pari}{]\! ]}

\def\>{\rangle}
\def\<{\langle}

\begin{document}
\renewcommand{\PaperNumber}{087}

\FirstPageHeading

\renewcommand{\thefootnote}{$\star$}

\ShortArticleName{Para-Grassmann Variables and Coherent States}

\ArticleName{Para-Grassmann Variables and Coherent
States\footnote{This paper is a contribution to the Proceedings of
the O'Raifeartaigh Symposium on Non-Perturbative and Symmetry
Methods in Field Theory
 (June 22--24, 2006, Budapest, Hungary).
The full collection is available at
\href{http://www.emis.de/journals/SIGMA/LOR2006.html}{http://www.emis.de/journals/SIGMA/LOR2006.html}}}

\Author{Daniel C. CABRA~$^{\dag^1\dag^2\dag^3}$, Enrique F.
MORENO~$^{\dag^3\dag^4}$ and Adrian TANAS\u{A}~$^{\dag^5}$}

\AuthorNameForHeading{D.C. Cabra, E.F. Moreno and A. Tanas\u{a}}

\Address{$^{\dag^1}$~Laboratoire de Physique Th\'eorique, CNRS UMR
7085,
Universit\'e L. Pasteur, \\
$\phantom{^{\dag^1}}$~3 rue de l'Uni\-ver\-sit\'e, F-67084 Strasbourg
Cedex, France}
\EmailDD{\href{mailto:cabra@lpt1.u-strasbg.fr}{cabra@lpt1.u-strasbg.fr}}

\Address{$^{\dag^2}$~Facultad de Inginer\'ia, Universidad Nacional de lomas de Zamora,\\
$\phantom{^{\dag^2}}$~Cno. de Cintura y Juan XXIII, (1832) Lomas
de Zamora, Argentina}

\Address{$^{\dag^3}$~Departamento de F\'isica, Facultad de
Ciencias Exactas,
Universidad Nacional de La Plata,\\
$\phantom{^{\dag^3}}$~C.~C. 67, 1900 La Plata, Argentina}

\Address{$^{\dag^4}$~Department of Physics, West Virginia University, \\
$\phantom{^{\dag^4}}$~Morgantown, West Virginia 26506-6315, USA}
\EmailDD{\href{Enrique.Moreno@mail.wvu.edu}{Enrique.Moreno@mail.wvu.edu}}

\Address{$^{\dag^5}$~Laboratoire de Physique Th\'eorique, B\^at.
210,
CNRS UMR 8627, Universit\'e Paris XI, \\
$\phantom{^{\dag^4}}$~F-91405 Orsay Cedex, France}
\EmailDD{\href{adrian.tanasa@ens-lyon.org}{adrian.tanasa@ens-lyon.org}}

\ArticleDates{Received September 29, 2006, in f\/inal form
November 22, 2006; Published online December 07, 2006}

\Abstract{The def\/initions of para-Grassmann variables and
$q$-oscillator algebras are recalled. Some new properties are
given. We then introduce appropriate coherent states as well as
their dual states. This allows us to obtain a formula for the
trace of a operator expressed as a function of the creation and
annihilation operators.}


\Classification{81R30; 81R50; 17B37}

\section{Introduction}

The study of dif\/ferent generalisations of Grassmann variables
and their applications has attracted a great deal of interest in
the last decades (see for example \cite{ohnuki, grec, baulieu,
para-alg1, para-alg2, chaichian, para-alg3, chai, q1, q2, rausch1,
kibler1, kibler, laplata}
and references therein).

Our approach is motivated by the fact that generalised Grassmann
variables provide a natural framework for the description of
particles obeying generalised statistics. We thus focus on the
$q$-oscillator algebra (introduced in \cite{q-boson1, q-boson2})
which is particularly appealing for our purpose for two distinct
reasons. First, the nilpotency property  of the creation and
annihilation operator is in direct connection with the maximal
occupation number of the studied particles. Second, for the case
of multi-particle states, the wave function acquires a nontrivial
phase when two particles are interchanged (one may recall that
this phase is trivial in the case of bosons and is $-1$ in the
case of fermions). One can note also that in \cite{altii} the
authors have discussed  many-body states and the algebra of
creation and annihilation operators for particles obeying
exclusion statistics.

This paper is structured as follows. We f\/irst review the
def\/inition and basic properties of the para-Grassmann variables.
We then re-examine the $q$-oscillator algebra and introduce
appropriate coherent states. New properties of the coherent states
are given. Finally we f\/ind a~representation for the trace of any
operator, as an integration over para-Grassmann variables. We show
that the trace can be represented as a para-Grassmann integral of
the matrix element of the respective operator on the coherent
state. This result is the natural generalisation of the usual
formula for the trace of an operator in the case of bosons or
fermions (see for example~\cite{book}). In the last section, some
perspectives are brief\/ly outlined. Let us mention here that this
work presents some partial results of a future
publication~\cite{noi}.

\section{One-particle states}

\subsection{Para-Grassmann variables}

Consider the non-commutative variables $\theta$ and $\bar \theta$:
\begin{gather} \label{def}
\theta^{p+1}=0,\qquad \bar \theta^{p+1} =0,\qquad \theta\bar
\theta = q^2 \bar \theta \theta, \qquad \mbox{where} \qquad
q^2=e^\frac{2\pi i}{p+1} %
\end{gather}
with $p$ some non-zero integer number. Note that in \cite{baulieu}
these variables are referred to as {\it classical
$(p+1)$-variables}. Moreover, the name ``para-Grassmann'' was used
also for dif\/ferent other def\/initions, see for example
\cite{ohnuki}, where some dif\/ferent variables, in connection
with para-statistics,  were def\/ined. Finally let us mention that
in \cite{q1,q2}, $q$-deformed classical variables and dif\/ferent
techniques were introduced.

We will use here the conventions of \cite{baulieu} for the
def\/initions of a dif\/ferential and integral calculus
appropriate for these variables. Thus \cite{baulieu,muie}
\begin{gather*}
\int  d\theta \, \theta^n=\delta^n_{p}\sqrt{[p]!},
\end{gather*}
where \begin{gather} \label{def-numar} [X]=\frac{q^{2X}-1}{q^2-1}
\end{gather} for any given $c$-number or operator $X$ and
\begin{gather*}
[n]!=[1]\cdots [n]
\end{gather*}
for any given number $n$. Of special importance is the {\it
q-deformed exponential}
\begin{gather*}
e_q^x = \sum\limits_{n=0}^{p} \frac{x^n}{[n]!} . %
\end{gather*}

\subsection[$q$-oscillator algebra]{$\boldsymbol{q}$-oscillator algebra}

Consider the {\bf $\boldsymbol{q}$-boson oscillator}
\cite{q-boson1, q-boson2}
\begin{gather} \label{q-boson} %
aa^\dag - q a^\dag a = q^{-N} ,\qquad aa^\dag - q^{-1} a^\dag a =
q^N,
\end{gather}
where $q\ne - 1$ is a complex number. Note that we deal here with
some generalisation of {\it bosons} and not of {\it fermions}.

Following the conventions of \cite{baulieu} we def\/ine
\begin{gather*}
\pard X \pari= \frac{q^X- q^{-X}}{q-q^{-1}},
\end{gather*}
for any $c$-number or operator $X$. If $q$ is a unit phase,
\begin{gather*}
\pard X \pari = \pard X \pari^*,
\end{gather*}
where by $^*$ we understand complex conjugation. We have
\[
a^\dag a = \pard N\pari.
\]
In particular, if $q^2=e^\frac{2\pi i}{p+1}$ we can write
\[
N= \frac{p+1}{\pi} \, \arcsin \, \left(a^\dag a \, \sin\,
\frac{\pi}{p+1}\right).
\]

\noindent {\it Occupation number representation:} Introducing a
vacuum vector $|0\rangle $ (s.t.\ $a|0\rangle =0$) we def\/ine
\begin{gather*} 
|n\rangle =\frac{(a^\dag)^n}{\sqrt{\pard n\pari!} }|0\rangle .
\end{gather*}
Using again the commutation relations \eqref{q-boson} we get
\begin{gather*} 
N|n\rangle = n |n\rangle,\qquad
a|n\rangle = \sqrt{\pard n \pari} |n-1\rangle,\qquad
a^\dag |n\rangle  =\sqrt{\pard n+1 \pari}|n+1\rangle .
\end{gather*}
Furthermore
\begin{gather} \label{in-rep} %
[N,a]=-a,\qquad
[N,a^\dag]=a^\dag
\end{gather}
and if $\,q=e^\frac{2\pi i}{p+1}\,$ it can be proved that the
creation and annihilation operators are $(p+1)$-nilpotent,
$\,a^{p+1}=0=(a^\dag)^{p+1}$ (see \cite{grec}).
Moreover, using \eqref{in-rep} we have that   for any
$c$-number~$\lambda$
\begin{gather*}
q^{\lambda N} a^\dag = q^\lambda a^\dag q^{\lambda N},\qquad
q^{\lambda N} a = q^{-\lambda} a q^{\lambda N}.
\end{gather*}
(see \cite{baulieu}).

If instead, we assume the nilpotency condition of creation and
annihilation operators
\begin{gather}
\label{nil}
 a^{p+1}=0=(a^\dag)^{p+1}.
\end{gather}
without imposing any condition on $q$ (here we require that the
exponent $p+1$ is the minimal exponent of nilpotency, so $a^r\neq
0\,$ and  $(a^{\dagger})^r\neq 0$ if $r \leq p$), we get
\[
a (a^\dag)^i = (1+ q^2+ \cdots + (q^2)^{i-1})q^{-N} (a^\dag)^{i-1}
+ q^i (a^\dag)^{i}a.
\]
If $q^2\ne 1$ this becomes
\[
a (a^\dag)^i = \frac{1-(q^2)^{i}}{1-q^2} q^{-N} (a^\dag)^{i-1} +
q^i (a^\dag)^{i}a.
\]
Taking now $i=p+1$ one has
\[
a (a^\dag)^{p+1} = \frac{1-(q^2)^{p+1}}{1-q^2} q^{-N} (a^\dag)^{p}
+ q^{p+1} (a^\dag)^{p+1}a.
\]
Now, using \eqref{nil} we derive that  $(q^2)^{p+1}=1$.

From the discussion above we conclude that for the $q$-boson
oscillator algebra \eqref{q-boson} the conditions: $q^2$  is a
primitive $p+1$ root of unity and  $a$ and $a^\dag$ are
$(p+1)$-nilpotent, are equivalent.

Finally, let us mention here that the operators $a$ and $a^\dag$
are hermitian conjugates and they generate a unitary
representation \cite{math}.

Consider now the change of variables
\begin{gather*}
b=q^\frac N2 a,\qquad \bar b = a^\dag q^\frac N2.
\end{gather*}
In these new variables the relations \eqref{q-boson} reads
\begin{gather} \label{not-boson} b\bar b - q^2 \bar b b =1,\qquad
b\bar b - \bar b b  =q^{2N}, \end{gather} and thus
\[
\bar b b = [N],
\]
where  $[N]$ was def\/ined in \eqref{def-numar}.

We can also express the occupation number states in terms of $\bar
b$ and $b$ as follows
\begin{gather*} 
|n\rangle  = \frac{(\bar b)^n}{\sqrt{[n]!}} |0\rangle ,\qquad
b|n\rangle =\sqrt{[n]} |n-1\rangle ,\qquad
\bar b |n\rangle = \sqrt{[n+1]} |n+1\rangle .
\end{gather*}
Furthermore
\[
[N,b]=-b,\qquad [N,\bar b]=\bar b.
\]

Unlike the operators $a$ and $a^\dag$, the operators $b$ and $\bar
b$ are not hermitian conjugates ($b^\dag\ne \bar b$) so in order
to def\/ine the dual vectors we introduce the operators $b^\dag$
and $\bar b^\dag$, the hermitian conjugate of $b$ and $\bar b$
respectively. One has (see also \cite{chai})
\begin{gather*} 
b^\dag = \bar b q^{-N}, \qquad \bar b^\dag= q^{-N} b.
\end{gather*}
Thus, up to a phase, $\bar b$ coincides with $b^\dag$ and $b$ with
$\bar b^\dag$. We then have
\begin{gather*}
\langle n|=\langle 0|\frac{b^n}{\sqrt{[n]!}},\qquad
\langle n|b=\langle n+1|\sqrt{[n+1]},\qquad
\langle n|\bar b = \langle n-1|\sqrt{[n]}.
\end{gather*}

Before going further let us mention that dif\/ferent $q$-deformed
algebraic structures with similar properties exist in the
literature, like  the para-Grassmann algebra (see \cite{para-alg1,
para-alg2, para-alg3})  or  the quon-algebra (see \cite{quon1,
quon2} and references therein).

 {\it Commutation relations between para-Grassmannians
and creation/annihilation operators.} We complete the set of
commutation relations given in equations \eqref{def} and
\eqref{not-boson} with
\begin{gather}
\label{set} \theta b = q^2 b \theta,\qquad
\theta \bar b = q^{-2} \bar b \theta,\qquad
\bar \theta \, \bar b = q^2 \bar b \, \bar \theta,\qquad
\bar \theta b = q^{-2} b \bar \theta
\end{gather}
(notice that instead of the set \eqref{set}, in some papers
\cite{kibler} regular commutation relations are assumed).

Thus, the structure we study further on consists of the nilpotent
operators $b$ and $\bar b$ obeying  the $q$-boson algebra
\eqref{not-boson}, and the para-Grassmann variables $\theta$ and
$\bar \theta$ obeying the commutation relations given in equations
\eqref{def} and \eqref{set}. We also set the value of $q$ to
$q=e^{\frac{\pi i}{p+1}}$.
Notice that, because of the commutation relations \eqref{set}, the
vectors $|n\rangle\,$ do not commute with $\theta$. Indeed, if we
impose $\theta |0\rangle =|0\rangle \theta$ it follows that
$\theta |n\rangle =q^{-2n}|n\rangle \theta$.

 {\it Coherent states.}
To f\/ind a coherent state $|\theta \rangle $ we write generically
\[
|\theta \rangle = \sum_{n=0}^p c_n |n\rangle.
\]
Imposing now
\begin{gather*} 
b|\theta\rangle =\theta |\theta\rangle
\end{gather*}
we get
\begin{gather*} 
|\theta\rangle =\sum_{n=0}^p \frac{q^{n(n+1)}}{\sqrt{[n]!}}
\theta^n |n\rangle
\end{gather*}
which can be written as
\begin{gather*} 
|\theta\rangle =e_q^{\bar b \theta}|0\rangle .
\end{gather*}
The action of $\bar b$ over the state $\theta$ is given by
\begin{gather*}
\bar b |\theta \rangle = q^{-2}\sum_{n=1}^p
q^{n(n+1)}\frac{[n]}{\sqrt{[n]!}} \theta^{n-1} |n\rangle .
\end{gather*}
Finally one has the scalar product
\begin{gather}
\label{scalar1}
 \langle n|\theta\rangle = \frac{q^{-n(n-1)}}{\sqrt{[n]!}} \theta^n.
\end{gather}

In analogy with the def\/inition of $|\theta\rangle $ we def\/ine
a dual state $\langle \bar \theta|$ through the relation
\begin{gather*} 
\langle \bar \theta| \bar b = \langle \bar \theta| \bar \theta.
\end{gather*}
We have
\begin{gather*} \langle \bar \theta |= \sum_{n=0}^p
\frac{q^{n(n-1)}}{\sqrt{[n]!}} \bar \theta^n\langle n| %
\end{gather*}
or, expressed in terms of~$b$,
\begin{gather*} \langle \bar \theta|=\langle 0|e_q^{\bar \theta b}.
\end{gather*}
Finally, we can compute the scalar product:
\begin{gather} \label{scalar2} \langle \bar \theta|n\rangle =
\frac{q^{n(n-1)}}{\sqrt{[n]!}}\, \bar\theta^n. \end{gather} Let us
stress that the scalar product \eqref{scalar2} {\it is not} the
complex conjugate of the scalar product~\eqref{scalar1}. First,
the para-Grassmannians $\theta $ and $\bar \theta$ cannot be
complex conjugated to each other (this is incompatible with the
commutation relations \eqref{def}) and second, $[n]$ is not a real
number.

Let us mention here that dif\/ferent def\/initions of coherent
states have been proposed for dif\/ferent algebraic structures in
some of the references cited. Thus, the def\/inition we give is
dif\/ferent by some phase (see equations \eqref{scalar1} and
\eqref{scalar2}) of the one proposed in \cite{baulieu} (also for
the $q$-boson oscillator algebra). Another example is given by the
def\/inition of \cite{kibler1, kibler}; here also the analytical
dif\/ference is given by some phase, but in \cite{kibler1, kibler}
the coherent states are def\/ined for a~dif\/ferent algebraic
structure.

The matrix elements of the identity operator can be written as
\begin{gather*} \langle \bar \theta| \, \text{Id} \, | \theta \rangle  =
\langle \bar \theta | \theta \rangle  = \sum_{n=0}^p
\frac{{1}}{{[n]!}} \, \bar \theta^n\theta^n.
\end{gather*} %
We can compute explicitly the matrix element $\,\langle \bar
\theta| \cal O | \theta\rangle  $ for any operator $\cal O$
expressed as a function of $\,b$ and $\,\bar b$. If in the case of
bosons and fermions this matrix element has a compact form,
independent of the form of $\cal O$ (see for example \cite{book}),
this does not hold anymore for para-Grassmannians.

Let us now look for a resolution of the identity
\begin{gather} \label{interm1} {\rm Id} =  \int d\bar \theta d\theta \mu
(\bar \theta, \theta) |\theta \rangle \langle \bar \theta|,
\end{gather}
where $\mu (\bar \theta, \theta)= \sum\limits_{n=0}^p  \mu_n \bar
\theta^n\theta^n $ is a weight factor to be determined ($\mu_n$
being some complex number coef\/f\/icient). Equation
\eqref{interm1} is equivalent to
\begin{gather} \label{interm2}
\delta_{mn}=\langle n|m\rangle = \langle n|\, {\rm Id}\, |
m\rangle = \int d\bar \theta d\theta \mu (\bar \theta, \theta)
\langle n|\theta \rangle \langle \bar \theta|m\rangle = \int d\bar
\theta d\theta \, \mu (\bar \theta, \theta) \frac{\theta^n \bar
\theta^m}{\sqrt{[n]![m]!}},
\end{gather}
where we have used expressions \eqref{scalar1} and \eqref{scalar2}
for the scalar products $ \langle n|\theta \rangle $ and $\langle
\bar \theta|m\rangle $. Notice that the $q$-factors involved in
these scalar products cancel each other, also since $\mu(\bar
\theta,\theta)$ only involves powers of $\bar \theta \theta$, it
commutes with $\langle n|$.

Integrating \eqref{interm2} we get (see \cite{baulieu})
\[
 \mu_n=\frac{(-1)^n}{[n]!} q^{n(n-1)}
 \]
so we f\/inally obtain
\[ \mu (\bar \theta, \theta) = \sum_{n=0}^p\frac{(-1)^n}{[n]!}
q^{n(n-1)} \bar \theta^n \theta^n = \sum_{n=0}^p
\frac{(-1)^n}{[n]!} \left( \bar \theta \theta\right)^n =
e_q^{-\bar \theta\theta}.
\]
Hence, we have the following resolution of the identity
\[ 
{\rm Id} = \int d\bar \theta d\theta \, e_q^{-\bar
\theta\theta}|\theta  \rangle \langle \bar \theta|
\]
thus allowing us to check the def\/inition of a coherent state
(see for example \cite{book2}).

 {\it Trace of an operator.}
Let us consider an operator $\cal O$ expressed as a function of
$b\,$ and $\,\bar b$. We want to express its trace in the form
\begin{gather} \label{tr1} {\rm Tr}\, {\cal O}=\int d\bar \theta d\theta
\rho (\theta, \bar \theta) \langle \bar \theta|{\cal
O}|\theta\rangle \end{gather}
with  $\rho (\theta, \bar \theta)$ some function to be determined.
We propose the following ansatz (that we will justify later)
\begin{gather}
\label{anstazrho} \rho (\theta, \bar \theta)= \sum_{n=0}^{p}\rho_n
\theta^n \bar \theta^n.
\end{gather}
Equation \eqref{tr1} can be written as
\begin{gather*}
 {\rm Tr}\, {\cal O}=\sum_{m,n=0}^p \int d\bar \theta
d\theta \rho (\theta, \bar \theta)  \langle \bar \theta|n\rangle
\langle n|  {\cal O} |m\rangle
\langle m|\theta\rangle \nonumber\\
\phantom{{\rm Tr}\, {\cal O}}{}=\sum_{m,n=0}^p \langle n|  {\cal
O} |m\rangle \int d\bar \theta d\theta \rho (\theta, \bar \theta)
\langle \bar \theta|n\rangle \langle m|\theta\rangle
\end{gather*}
so we have
\[
\int d\bar \theta d\theta \rho (\theta, \bar \theta)  \langle \bar
\theta|n\rangle \langle m|\theta\rangle = \delta_{nm}.
\]
Since only terms with $n=m$ are nonzero, the function
$\rho(\theta, \bar\theta)$ can only have terms with the same
powers of $\theta$ and $\bar \theta$, in agreement with our ansatz
\eqref{anstazrho}. A straightforward computation gives
\[
 \rho_n= \frac{(-1)^n}{[n]!}q^{(n+1)(n+2)}
 \]
so we get
\begin{gather} \label{trace1} {\rm Tr}\, {\cal O}=\int d\bar \theta d\theta
\sum_{n=0}^{p} \frac{(-1)^n}{[n]!}q^{(n+1)(n+2)} \theta^n \bar
\theta^n \langle \bar \theta|{\cal O}|\theta\rangle . \end{gather}
(In the framework of the para-Grassmann algebra mentioned above,
some related calculations have been performed in
\cite{para-alg3}.)

The importance of formula \eqref{trace1} comes from the fact that
it is a direct generalisation of  the trace formula for boson and
fermion coherent states (see for example \cite{book}). Following
the same line of reasoning it is more useful to use this formula
rather than the trace expressed in terms of occupation states for
the computation of some specif\/ic quantities (like for example
the partition function or the occupation number). Furthermore,
this would represent a direct generalisation of the calculations
performed in the case of bosons or fermions.

\section{Perspectives}

In this paper we have studied para-Grassmann variables and the
$q$-oscillator boson algebra. Appropriate coherent states were
def\/ined and some new properties studied. Finally we obtained
a~trace formula for any operator $\cal O$ expressed as a function
of the creation and annihilation operators. This formula is
expressed as an integral over para-Grassmann variables of the
coherent-state matrix elements of the operator $\cal O$.

The next step in the direction of work we propose here is to
generalise these results to multi-particle states. Once one has
the equivalent of the trace formula \eqref{trace1} for
multi-particle states, one can calculate several physical
quantities, like the partition function and occupation number. The
results can then be compared with the behaviour of particles
obeying generalised statistics. We will report on these issues in
a future paper \cite{noi}.

\subsection*{Acknowledgments}

A.~Tanas\u{a} acknowledges M.~Goze and R.~Santachiara.

\LastPageEnding

\end{document}